\documentclass[nofootinbib,apsspec,showpacs,twocolumn,groupedaddress,lengthcheck,preprintnumbers]{revtex4}
\usepackage{amsmath}
\usepackage{amssymb}
\usepackage{graphics}
\usepackage{slashed}
\usepackage[dvips]{graphicx}
\usepackage{graphicx}
\usepackage[usenames]{color}

\def\simge{\mathrel{%
       \rlap{\raise 0.511ex \hbox{$>$}}{\lower 0.511ex \hbox{$\sim$}}}}
\def\simle{\mathrel{
       \rlap{\raise 0.511ex \hbox{$<$}}{\lower 0.511ex \hbox{$\sim$}}}}

\newcommand \beq{\begin{eqnarray}}
\newcommand \eeq{\end{eqnarray}}

\usepackage[colorlinks=true,linktocpage=true,linkcolor=blue,citecolor=blue]{hyperref}

\begin{document}
\title{Understanding the puzzle of angular momentum conservation\\ in beta decay and related processes
}
\author{Gordon Baym,$^{a,b}\footnote{Corresponding author}$ Jen-Chieh Peng$^a$ and C.\ J.\  Pethick$^{a,b,c}$}
\affiliation{\mbox{$^a$Department of Physics, University of Illinois, 1110
  W. Green Street, Urbana, IL 61801} \\
\mbox{$^b$The Niels Bohr International Academy, The Niels Bohr Institute, University of 
Copenhagen,}\\
\mbox{Blegdamsvej 17, DK-2100 Copenhagen \O, Denmark}\\	
\mbox{$^c$NORDITA, KTH Royal Institute of Technology and Stockholm University}\\ 
\mbox{ Hannes Alfv\'ens v\"ag 12, SE-10691 Stockholm, Sweden} 
}

\date{\today}

\begin{abstract}
     We ask the question of how angular momentum is conserved in electroweak interaction processes.  To introduce the problem with a minimum of mathematics, we first raise the same issue in elastic scattering of a circularly polarized photon by an atom, where the scattered photon has a different spin direction than the original photon, and note its presence in scattering of a fully relativistic spin-1/2 particle  by a central potential.    We then consider inverse beta decay in which an electron is emitted following the capture of a neutrino on a nucleus.    While both the incident neutrino and final electron spins are antiparallel to their momenta, the final spin is in a different direction than that of the neutrino -- an apparent change of angular momentum.   However, prior to measurement of the final particle, in all these cases angular momentum is indeed conserved,    The apparent non-conservation of angular momentum arises in the quantum measurement process in which the measuring apparatus does not have an initially well-defined angular momentum, but is localized in the outside world.    We generalize the discussion to massive neutrinos and electrons, and examine nuclear beta decay and electron-positron annihilation processes through the same lens, enabling physically transparent derivations of angular and helicity distributions in these reactions.  \\

\end{abstract}

\maketitle

\section{Introduction}

  How is angular momentum conserved in simple processes such as elastic scattering of a photon from an atom. scattering of a relativistic particle with spin from a central potential, or in 
beta and inverse beta decays?    In elastic scattering of a right circularly polarized photon on a heavy atom in its
ground state, at 90$^\circ$  the photon emerges with linear polarization transverse to the reaction plane defined by
the initial and final photon directions.  In potential scattering of a relativistic spin-1/2 particle, the spin of the particle is rotated towards the direction in which the particle propagates after the scattering, and in the ultrarelativistic limit, the helicity of the particle is conserved \cite{Land L}.     Or, consider 
inverse beta decay in which a massless antineutrino propagating in the z direction interacts with a nucleus, leading to a final massless  (for the sake of the argument) electron propagating at an angle $\theta$ with respect to the
z axis.  If we focus only on a Fermi process with no spin transfer to the nucleus, the final nuclear spin state is the same as the initial spin state.  The incident neutrino has spin -1/2 along the z axis, but the final electron has spin -1/2 along the tilted axis.
What happens in all these cases to the original spin angular momentum of photon, the relativistic particle or the neutrino?    The problem is that the initial and final spin angular momenta are seemingly not equal in these examples; how is angular momentum conserved?    

  One might argue that orbital angular momentum must be transferred to the atom, the scattering center, or the nucleus to conserve the total angular momentum.  In the atomic case, when the atom is initially and finally in its internal ground state, any transfer of orbital angular momentum must be in the direction orthogonal to the reaction plane, which does not help.  In potential scattering of a relativistic particle, the rotation of the spin occurs even if the scattering center is infinitely massive and has no kinetic degrees of freedom.  If we consider, in inverse beta decay, a superallowed transition in which the initial and final nuclei have spin 0,\footnote{The physics is clearest for a superallowed transition, from an initial $0^+$ state to a final $0^+$, since one need not worry about the angular momentum absorbed by the nucleus.   A simple example is the reaction
  $\bar \nu_e + ^{14}$O$ \to e^+ + ^{14}$N.}   then the only possibility would be a transfer of angular momentum to the center of mass of the nucleus.   However, the weak interaction has a range of order the inverse of the mass of the W boson, $\hbar/m_W c$ (where $c$ is the speed of light), and thus for incident neutrino energies $E_\nu \ll m_Wc^2$ there is little amplitude (a function of $E_\nu/m_Wc^2$) to have anything but zero orbital angular momentum transfer in the interaction.   In the following we take units, $\hbar=c=1$.

     In these examples, the angular momentum structure of the final state is constrained, for photons by the condition that the electromagnetic field of the emitted photon must be transverse to the direction of motion of the photon, and in the inverse beta decay by the condition that the final electron must be in a negative chirality eigenstate, which for massless electrons, requires the electron spin to be antiparallel to the electron direction.   Even potential scattering of massless particles is constrained by the same condition of conservation of chirality.  
   
  One can see the same problem even in the absence of a central scatterer by considering, for example, the annihilation reaction of an electron antineutrino-neutrino pair, $\bar\nu_e +  \nu_e \to \bar\nu_\mu +  \nu_\mu$, into a muon antineutrino-neutrino pair.  The two initial neutrinos, assumed massless, together have spin angular momentum unity along the direction of the $\bar\nu_e$, and the muon neutrinos, emerging at an angle $\theta$ with respect to the initial electron neutrino axis, have spin angular momentum unity along the direction of the $\bar\nu_\mu$.  Again, the spin angular momentum is rotated by angle $\theta$, raising the question of  how angular momentum is conserved in this process.
    
   The key is that in none of the cases considered above is the final particle (or particles) actually emitted in a momentum  eigenstate.  Rather in all cases,  the particle is emitted in a spherical wave which conserves the initial angular momentum, as we work out in detail below.  Only by making a measurement does one determine that the final particle is emitted in a given direction, and only   after the measurement does angular momentum appear not to be conserved.    Understanding conservation of angular momentum when a particle is detected propagating in a particular direction requires that one consider transfer of angular momentum to the quantum measuring apparatus.  The
apparent non-conservation of angular momentum involves similar issues in a Stern--Gerlach measurement; see Refs.~\cite{araki,sahel} for discussions.

        To illustrate the basic physics, we first write out, in Sec.~\ref{rayleigh}, the detailed state of a photon after elastic scattering from 
an atom, showing that the photon angular momentum resides not simply in the spin of the photon but also in its orbital angular momentum.   In Sec.~\ref{spherical} we construct the analogous electron spherical wave solution for massless electrons in inverse beta decay, and show that the (s-wave) scattering of a massless spin-1/2 particle from a central potential has an identical structure.        Then in Sec.~\ref{gamma-hel}  we describe the modification of these results for massive neutrinos or antineutrinos taking part in a weak interaction.   We turn to inverse beta decay in Sec.~\ref{inverse}, which is simpler to analyze than beta decay, which we consider in Sec.~\ref{betadecay}.    We discuss in Sec.~\ref{pairs} the similar issues with angular momentum appearing in lepton pair annihilation processes at high energies, e.g., $e^+ + e^- \to \mu^+ + \mu^-$.  As we see throughout, working in terms of the angular momentum state prior to measurement yields physical insights into these various processes. We conclude in Sec.~\ref{conclusion}.

 \section{Elastic scattering of light}
 \label{rayleigh}
      
  We look now at the photon scattering problem in more detail, considering a right circularly polarized photon incident along the z axis being scattered elastically by an atom which remains in its ground state (angular momentum $J=0$).   This problem is basically Rayleigh scattering, with the specifics of the internal states of the atom playing no essential role.    Crucially, the scattering transfers photon spin to photon orbital angular momentum, conserving the total photon angular momentum.\footnote{Author GB is grateful to Immanuel  Bloch for raising, at the Aspen Center for Physics, the importance of the photon orbital angular momentum in light scattering by an atom.}
     
    The orbital angular momentum of the scattered photon is seen simply in the dipole approximation, where the incoming right circularly polarized photon induces a dipole moment $ \vec d \sim (1,i,0)^T$, in Cartesian coordinates.
The electric field, $\vec{\cal E}$, of the re-emitted photon is proportional to the component of $\vec d$ transverse to the angular direction $\hat k$ of the photon,  
$\vec{\cal E} \sim \vec d - \hat k (\hat k\cdot \vec d)$, and the re-emitted photon is described by a vector wave function, $\Psi_\gamma$, proportional to $\vec{\cal E}$.  
Since $\hat k = (\sin\theta \cos\phi, \sin\theta \sin\phi, \cos\theta)^T$, where $T$ denotes the transpose, and $\vec d\cdot \hat k \sim \sin\theta e^{i\phi}$, the angular dependence of $\Psi_\gamma$ is given by
\beq
  \Psi_\gamma(\theta,\phi)  &\sim& \vec{\cal E} \sim (1-\frac12\sin^2\theta) (1,i,0)^T \nonumber\\ && \hspace{-24pt} - \frac12\sin^2\theta e^{2i\phi}(1,-i,0)^T\ - \cos\theta\sin\theta e^{i\phi} (0,0,1)^T; \nonumber\\
  \label{decomp}
 \eeq
 the three terms in $\Psi_\gamma$ have photon spin, $S_z$, and orbital angular momentum, $L_z=-i\partial/\partial \phi$, given by  $(S_z,L_z) = (1,0), (-1,2)$, and (0,1), respectively.   The first and third terms contain total orbital angular momentum components $L$ = 0 and 2, while the middle term has  $L$ =  2.  
 The final electromagnetic field has the same total angular momentum $J = 1, J_z = 1$ as the original photon, but its intensity varies with direction.   At 90$^\circ$, the final term in Eq.~\eqref{decomp} drops out, and the photon state is a linear combination of a term with the photon right circularly polarized along z, with $L_z=0$, and a second with the photon left circularly polarized along z.  This latter term raises the specter of nonconservation of angular momentum, until one realizes that it in fact carries orbital angular momentum, $L_z= +2$, so that this component has total $J_z=1$ as well.  

   Imagine now trying to detect the full angular momentum of the scattered photon by absorption by a detector atom located in a specific
direction to the scattering atom which is different from that of the incoming photon.   Since the orbital angular momentum carried by a photon depends on the quantum mechanical amplitude for all angles, a measurement at a single angle cannot give complete information about its orbital angular momentum.  The localization of the detector atom, if it is in a pure state, implies that it is in a superposition of states with different orbital angular momenta, or if it is in a mixed state, it is described by a density matrix with components having different orbital momenta.
Nonetheless,  the internal transitions of the detector atom are governed by the spin angular momenta of the two components of the plane polarized photon, and thus such a measurement can determine only the spin angular momenta of the components of the scattered photon.

\section{Elastic scattering of massless spin-1/2 particles. }
\label{spherical}

    The analyses of the scattering of relativistic spin-1/2 particles and of inverse beta decay are similar to that of the scattered photon above, complicated only by the Dirac technology describing the chirality of the scattered fermion.  We look first at inverse beta decay,  which as we argue can be regarded as a particular case of scattering of a massless spin-1/2 particle.  To simplify the discussion we consider only Fermi  beta decay processes, and neglect the recoil of the nucleus involved.  We first construct the wave function for massless electrons and neutrinos, and in Sec.~\ref{inverse}  generalize the result to non-zero electron and neutrino masses. 

    The incident neutrino is described by a Dirac spinor of the form,
\beq
  \Psi_\nu(r) \sim e^{i\vec p\cdot \vec r} (0,1,0,-1)^T.
  \label{incident}
\eeq
 The electron emerges with the same angular momentum as the incident neutrino.  We write the Dirac spinor for the electron emerging from the inverse beta decay in terms of its upper and lower components as
\beq
     \Psi(\vec r\,) = (\varphi,\chi)^T.
\eeq   
The upper component s-wave solution of the Dirac equation for the electron has the form
\beq
  \varphi =   h_0(pr)(0,1)^T,
  \label{upper}
\eeq
with $p = \sqrt{E^2-m^2} \to E$, and
$h_0(x) = ie^{ix}/x$ the lowest spherical Hankel function.     Similarly the lower two components have the general form,
\beq
 \chi &=&\frac{1}{ip }\vec \tau\cdot \vec\nabla \varphi, 
 \eeq
 where the $\tau$ are the Pauli matrices.
 We assume always that $r\gg 1/p$, where $h_0' = i h_0$ and therefore
 \beq
 \chi &= & h_0(pr) (\sin\theta e^{-i\phi }, -\cos\theta)^T, 
 \label{lower}
\eeq
where $\theta$ and $\phi$ are the usual polar and azimuthal angles measured with respect to the incident neutrino direction.   The emergent electron state for a contact interaction prior to projecting onto negative chirality (denoted by the superscript 0) is a spherical wave, 
 \beq
\Psi^0_{ed}(\vec r)  = h_0(pr)\big(0,1 , \sin\theta e^{-i\phi},- \cos\theta\big)^T,
 \label{Le}
\eeq
with total angular momentum $(J,J_z)=(1/2,-1/2)$ measured with respect to the neutrino direction.    The spinor in this state is in fact that of
a massless particle of spin down along the z direction, as denoted by the subscript $d$, and momentum $p$ in the $(\theta,\phi)$ direction.

However, the $1-\gamma_5$ in the weak interaction requires projecting this state along $\gamma_5=-1$, with the operator $(1-\gamma_5)/2$,  to give the final electron state:
\beq
   &&   \Psi_{ed}(\vec r\,) = \frac{h_0(pr)}{2}\big(-\sin\theta e^{-i\phi },1 + \cos\theta, \nonumber\\ && \hspace{72pt} \sin\theta e^{-i\phi },-(1 +\cos\theta)\big)^T.
 \label{fes0}
\eeq
The projection does not change the angular momentum, since $\gamma_5$ commutes with both the spin and orbital angular momentum operators.   This state is the analog of the photon wave $\Psi_\gamma$, Eq.~\eqref{decomp}.

  Since the operator $\vec\tau\cdot\vec\nabla$ is rotationally invariant, the angular momentum of the state $\Psi$ is simply that of $\varphi$.  The angular momentum of  $\varphi \sim (0,1)$  is clearly $J=1/2, J_z=-1/2$.    The state $\Psi_{ed}$ is neither an eigenstate of $S_z$ nor of $L_z$; explicitly, 
\beq
   S_z \Psi_{ed} &=& \nonumber\\ &&\hspace{-12pt} \frac{h_0}{4}(-\sin\theta e^{-i\phi },-1 - \cos\theta, \sin\theta e^{-i\phi }, 1 + \cos\theta)^T \nonumber\\  
   L_z\Psi_{ed} &=& -i\frac{\partial \Psi_{ed} }{\partial\phi}\nonumber\\ && \hspace{-12pt} = \frac{h_0}{2}(\sin\theta e^{-i\phi },0, -\sin\theta e^{-i\phi },0)^T.
\eeq   
However it is an eigenstate of $J_z=S_z+L_z$ with eigenvalue -1/2. 

  We can rewrite the state in \eqref{fes0} as
\beq
   &&   \Psi_{ed}(\vec r) =  \sqrt2 {h_0(pr)}e^{-i\phi/2}\cos(\theta/2){B}_L,
    \label{fes1}
\eeq
where the spinor, $B_L$ (normalized to unity), is given by
\beq
  {B}_L &=&e^{-i\phi\sigma_z/2} e^{-i\theta \sigma_y/2}(0,1,0,-1)^T\nonumber\\   
 &=&\frac1{\sqrt2}\big(-\sin(\theta/2)e^{-i\phi/2},\cos(\theta/2)e^{i\phi/2}, \nonumber\\ &&\hspace{30pt}\sin(\theta/2)e^{-i\phi/2},-\cos(\theta/2)e^{i\phi/2}\big)^T .
     \label{bl}
\eeq
 Although  $\Psi_{ed}(\vec r\,)$ is a spherical wave with angular momentum along the negative z direction, its value in a given direction $(\theta,\phi)$ describes a left-handed particle [spin 1/2 in the direction ($\pi-\theta, \phi+\pi$) opposite to ($\theta, \phi$)], as it must by chiral invariance.   
This state is a hedgehog solution;  its component along any given direction describes an electron with its spin pointing opposite to the direction.
It is analogous to a simple spin 1/2 along the z direction being a superposition of spin up and spin down states along the x direction.  The direction of the electron momentum and its spin are highly entangled in the spherical wave.\footnote{The requirement that an electron in inverse beta decay emerge from a weak interaction with negative chirality, and in scattering of massless fermions as well, imposes an effective spin-orbit coupling on the electron at its creation, strongly correlating its spin and momentum. To see this connection, we write $\gamma_5$, for arbitrary direction $\hat n$, as $\gamma_5= (\hat n \cdot \vec\alpha) (\hat n\cdot\vec\Sigma)$,
in terms of the Dirac operators $\alpha^i =  \gamma^0\gamma^i$ and $\Sigma_i = i\epsilon_{ijk} \gamma^j\gamma^k$, the Dirac spin operator. Since $\vec\alpha =\delta H_0/\delta \vec p$  is effectively the operator for the electron velocity, where  $H_0$ is the free Dirac Hamiltonian, the chiral projection operator, $(1-\gamma_5)/2$, has the explicit form of  a spin-orbit force.   We thank Wolfgang Ketterle for an incisive comment relating spin-orbit forces to the weak interaction problem.  Similarly the transversality condition in photon scattering is, in a sense, a spin-orbit coupling.}   

      The elastic (nominally s-wave) scattering of a massless spin-1/2 particle has exactly the same structure.      We assume that the particle of negligible mass $m$ is,  for simplicity an electron, incident along the z direction with momentum $p \gg m$, with spin $s_z=-1/2$, described by the same state, Eq.~\eqref{incident}, as the neutrino in inverse beta decay.  Importantly, this state not only has definite helicity,  $2\hat s\cdot \hat p =-1$, it is also a state of definite chirality, $\gamma_5 = - 1$.  Thus the final electron state must also have the same chirality, which is conserved for massless particles.   The final state is the same, Eq.~\eqref{fes0}, as that of the final electron in inverse beta decay.  The only difference between the scattering problem and inverse beta decay is that in the latter, the incident particle changes its species in interacting with the scattering nucleus.
      
     While in inverse beta decay with massless fermions, or in the scattering problem, the final electron emerges as a spherical wave,  $\Psi_{ed}(\vec r\,)$, with the angular momentum parallel to that of the incident fermion, after a measurement determining the direction of the electron its spin then becomes antiparallel to this direction.  An immediate question is how does the measurement of the final electron  direction lead to its angular momentum being apparently rotated from the initial angular momentum along the incident axis to that of the final particle?   
     
      The problem, when all is said and done, is that the measuring apparatus, e.g., a wire chamber, being localized in direction, cannot be  prepared with a well defined orbital angular momentum; neither can a spatially localized apparatus measure changes of the electron's angular momentum.   Being able to detect the electron direction precludes being able to detect the electron orbital angular momentum.    Even if one were to imagine a microscopic measuring apparatus originally in a well defined quantum state, the combined state of the electron and the measuring apparatus after its encounter with the electron would be entangled.   
 The wave function of the electron plus apparatus would be a sum of terms, each of which conserves total angular momentum, with the electron in a defined spin state and the total angular momentum  shared between the electron and apparatus.  As for the scattered photon above, without determining the orbital angular momentum of the apparatus one cannot see conservation of total angular momentum in play. 
  
   A related question of conservation of angular momentum arises in the gravitational spin-Hall effect, in which  a photon (or indeed other spinning object) passing by a gravitating body is lifted up out of the orbital plane if right circularly polarized, and is pushed down out of the plane if left circularly polarized \cite{LZ,bliokh,naoki,andersson,Noh};
the photon spin is not conserved in its motion, but rather changes direction, reminiscent of the apparent change in spin direction from the initial neutrino to the final electron in inverse beta decay.    The change of the spin angular momentum is brought about by an effective spin-orbit force which, unlike in inverse beta decay, transfers angular momentum from spin to orbital.   A further difference is that  in weak interactions at low energy the particle wavelengths are large compared with the range of the weak interaction (e.g., $p_\nu \ll m_W$),  while the spin-Hall effect arises as a first order correction to geodesic motion in the opposite limit that the wavelength of light is small compared with the characteristic scale height of the gravitational potential.  We note that relic neutrinos also undergo a spin-Hall effect in their propagation through the gravitational fields in the Universe \cite{nugrav}.

\section{Helicity and $\gamma_5$ eigenstates for finite mass}
\label{gamma-hel} 

 We now take into account the finite mass of neutrinos and electrons, and focus on the angular distribution of final particles in inverse and direct beta decay.   Finite mass neutrinos can have positive in addition to negative helicity.   The relative amplitudes for a weak vector current source
proportional to $1-\gamma_5$ to produce a neutrino of negative or positive helicity are $\sqrt{(1\pm \beta_\nu)/2}$, respectively ~\cite{mdbd}.

  To see these amplitudes we note that the Dirac spinor, normalized to unity, of a negative helicity neutrino (or antineutrino) of energy $E$, propagating in the +z direction, say, is
\beq
  u_{\nu L}  &=& \sqrt{\frac{E+m}{2E}}\left(0,1,0,-\frac{p}{E+m}\right)^T \nonumber\\ &=& (0,W_+,0,-W_-)^T, 
  \label{unuL}
 \eeq
where we take the spin quantization axis along +z,  and $W_\pm \equiv  \sqrt{(E\pm m)/2E}$.  Since  $W_+^2+W_-^2 = 1$ and $W_+W_-  = \beta_\nu/2$,
where $\beta_\nu = v_\nu/c$, with $v_\nu$ the neutrino velocity,
it follows that $(W_+\pm W_-)^2 = 1 \pm \beta_\nu$, and thus
\beq
 &&W_\pm  = \frac12\left( \sqrt{1+\beta_\nu}\pm\sqrt{1-\beta_\nu}\right). 
 \label{Wpm}
  \eeq
Similarly, the spinor of a positive helicity neutrino (or antineutrino) propagating in the +z-direction is 
 \beq
  u_{\nu R} &=& \sqrt{\frac{E+m}{2E}}\left(1,0,\frac{p}{E+m},0\right)^T  = (W_+,0,W_-,0)^T.
 \nonumber\\
 \eeq
  
   In the weak interaction, the projector  $\frac12(1-\gamma_5)$ acting on the  4-component neutrino creation operator is
\beq
&&\frac12(1-\gamma_5)\left[ u_{\nu L}a_{\nu L}^\dagger + u_{\nu R}a_{\nu R}^\dagger\right] \nonumber\\
&&=
  \sqrt{\frac{1+\beta_\nu}2} u_{\gamma_5=-1,d}a_{\nu L}^\dagger
   + \sqrt{\frac{1-\beta_\nu}2} u_{\gamma_5=-1,u}a_{\nu R}^\dagger, \nonumber\\
\eeq
showing the amplitudes $\sqrt{(1\pm \beta_\nu)/2}$ explicitly.  Here  the $a_{\nu L(R)}^\dagger$ create neutrinos in $L$ or $R$ states,   
and the $\gamma_5$ eigenstates (normalized to unity) for spin up or down (denoted by $u$ or $d$) along the $z$-direction are
\beq
u_{\gamma_5=\pm,u} &= &\frac{1}{\sqrt2}(1,0,\pm1,0)^T 
\,\,\,{\rm and}\nonumber\\
\quad u_{\gamma_5=\pm,d} &=&  \frac{1}{\sqrt2}(0,1,0,\pm1)^T.
\label{helgamma5}
\eeq  

The helicity eigenstates can be written in terms of the $\gamma_5$ eigenstates as
\beq
  u_{\nu L}  
  &=&   \sqrt{\frac{1+\beta_\nu}{2}}u_{- d}+\sqrt{\frac{1-\beta_\nu}{2}}u_{+ d}.
   \label{h-1gamma}
\eeq
and
\beq
  u_{\nu R}   &=&  \sqrt{\frac{1-\beta_\nu}{2}} u_{- u}+\sqrt{\frac{1+\beta_\nu}{2}}u_{+ u}.
  \label{h1gamma}
\eeq
The amplitude for a neutrino with left-handed helicity to be in a $\gamma_5=\mp 1$ eigenstate is therefore $\sqrt{(1\pm\beta_\nu)/2}$ and with right-handed helicity $\sqrt{(1\mp\beta_\nu)/2}$.  The expectation value $\langle h\rangle $ of the helicity of a neutrino made with $\gamma_5=-1$  is simply $-\beta_\nu$, and for an antineutrino with $\gamma_5=1$ is $\beta_{\bar\nu}$.  

  The same technology applies to finite mass fermions undergoing nominally s-wave scattering.

\section{Inverse beta decay}
\label{inverse}

  We now study the angular momentum in inverse beta decay for finite mass neutrinos and electrons.   (Many of the results of this section are explicit or implicit  in Ref.~\cite{long}, but the derivations here are more direct.) 

   Independent of the masses of the neutrino and electron, only the s-wave component of the initial neutrino wave packet (below GeV energies) is affected by the weak interaction, and emerges as the $\gamma_5= -1$ component of a spherical wave solution of the Dirac equation.   The only modification from Eqs.~\eqref{upper} and \eqref{lower} is that the upper and lower components of the solution of the Dirac equation for the incident neutrino having left-handed helicity is that  \beq
  \varphi = \sqrt2 V_+ h_0(pr)(0,1)^T,
\eeq
and 
\beq
 \chi &=&-\sqrt2\, i\frac{V_-}{p}\vec \tau\cdot \vec\nabla \varphi \nonumber\\ &&=  -\sqrt2 \,iV_-h_0'(pr) (\sin\theta e^{-i\phi }  -\cos\theta)^T, 
\eeq
with (cf. Eq.~\eqref{Wpm})
\beq
   V_\pm =  \frac12\left( \sqrt{1+\beta_e}\pm\sqrt{1-\beta_e}\right). 
\eeq
 
   Again the electron spin state is the same as that of the incident neutrino.  \ If the incident neutrino has left-handed helicity, the emergent electron state for a contact interaction without the $1-\gamma_5$ would be a spherical wave,
\beq
\Psi^0_{ed}(\vec r\,)  =\sqrt2 h_0(pr)\big(0,V_+ , V_-\sin\theta e^{-i\phi},- V_-\cos\theta\big)^T, \nonumber\\
 \label{Le}
\eeq
with angular momentum $(J,J_z)=(1/2,-1/2)$ measured with respect to the neutrino direction. Projection
 along $\gamma_5=-1$  (with $(1-\gamma_5)/2$) gives the final electron state, 
\beq
   &&   \Psi_{ed}(\vec r\,) =\frac{h_0(pr)}{\sqrt2}\big(-V_-\sin\theta e^{-i\phi },V_+ + V_-\cos\theta, \nonumber\\ && \hspace{60pt} V_-\sin\theta e^{-i\phi },-(V_+ + V_-\cos\theta)\big).
   \label{psied}
\eeq

Similarly, if the incident neutrino has right-handed helicity, the electron emerges in the state,
\beq
\Psi^0_{e u}(\vec r\,)= \sqrt2   h_0(pr)  \big(V_+, 0,  V_-\cos\theta, V_-\sin\theta e^{i\phi}\big)^T,
 \label{Re}
\eeq
prior to taking the  $\gamma_5=-1$ component, and after,
 \beq
    &&  \Psi_{eu}(\vec r\,) =\frac{h_0(pr)}{\sqrt2}\big(V_+ -V_-\cos\theta,  -V_-\sin\theta  e^{i\phi},  \nonumber\\ && \hspace{72pt}   -( V_+ -V_-\cos\theta),  V_-\sin\theta  e^{i\phi} \big)^T ,\nonumber\\
    \label{psieu}
\eeq
with angular momentum $(J,J_z)=(1/2,1/2)$, as denoted by the subscript $u$ (for up).   Again, $\Psi^0_{e u}(\vec r\,)$ is neither an eigenstate of  $S_z$ or $L_z$ individually.   Note that while $\Psi^0_{eu}$ and $\Psi^0_{ed}$  are orthogonal, $\Psi_{eu}$ and $\Psi_{ed}$ are not.

As earlier, we write the states $ \Psi_{ed}(\vec r\,)$ and $ \Psi_{eu}(\vec r\,)$ in terms of spin eigenstates along the direction $(\theta, \phi)$,
\beq
   &&   \Psi_{ed} = h_0(pr)\left(a^{(d)}_L  {B}_L +a^{(d)}_R {B}_R\right)  \,\,\,{\rm and} \nonumber\\
   &&   \Psi_{eu} = h_0(pr)\left(a^{(u)}_L  {B}_L +a^{(u)}_R {B}_R\right), 
  \label{BLBR}
 \eeq
with the amplitudes for the electron in  $\Psi_{ed}$ to be in a left- or right-handed helicity state, respectively, in the $(\theta,\phi)$ direction are
\beq
  a^{(d)}_L& =&  \sqrt{1+\beta_e}\cos(\theta/2)e^{-i\phi/2}, \nonumber\\  a^{(d)}_{R}&=& \sqrt{1-\beta_e}\sin(\theta/2)e^{-i\phi/2}, 
    \label{aL}
\eeq
and for an electron in $\Psi_{eu}$ 
\beq
  a^{(u)}_{L} &=& - \sqrt{1+\beta_e}\sin(\theta/2)e^{i\phi/2}, \nonumber\\ a^{(u)}_{R}&=& \sqrt{1-\beta_e}\cos(\theta/2)e^{i\phi/2}.
  \label{aR}
\eeq
 The spinor ${B}_L$,  Eq.~\eqref{bl}, is that for left-handed electron spin and the  spinor $B_R$, given by
 \beq
  {B}_R  &=&e^{-i\phi\sigma_z/2} e^{-i\theta \sigma_y/2}(1,0,-1,0)^T \nonumber\\   
    &=& \frac1{\sqrt2}\big(\cos(\theta/2)e^{-i\phi/2} ,\sin(\theta/2)e^{i\phi/2 }, \nonumber\\ &&\hspace{16pt}-\cos(\theta/2)e^{-i\phi/2} ,-\sin(\theta/2)e^{i\phi /2}\big)^T,
\eeq
is that for right-handed electron spin in the $(\theta,\phi)$ direction; both spinors are $\gamma_5=-1$ eigenstates, the analogs of the $u_{-,s}$ in Eq.~\eqref{helgamma5}.  

    The angular distribution of the electron in inverse beta decay, when the incident neutrino is left-handed,  is given by 
\beq
  |\Psi_{ed}(\vec r\,)|^2 \sim 1+\beta_e \cos\theta,
\eeq
while if the incident neutrino is right-handed, the electron emerges in with an angular distribution
\beq
  |\Psi_{eu}(\vec r\,)|^2 \sim 1-\beta_e \cos\theta.
\eeq
The right and left-handed neutrino states are not coherent, since the second fermion involved in the incident neutrino production, e.g., the electron in an initial beta decay, carries away phase information.
Including the probabilities of the handedness of the incident neutrino, we find the total angular distribution of the electron,
\beq
   \frac{d\sigma}{d\theta} &\sim& \frac{1+\beta_\nu}2(1+\beta_e \cos\theta) + \frac{1-\beta_\nu}2 (1-\beta_e \cos\theta) \nonumber\\
   &= & 1+\beta_e\beta_\nu \cos\theta.
   \label{eang}
\eeq

   To determine the expectation value of the electron spin, we assume that in an inverse beta decay the final electron is emitted in the (x,z) plane.   Its mean spin, $\langle \vec\Sigma\, \rangle$, as a function of $\theta$ in the states $\Psi_{ed}$ and $\Psi_{eu}$, is given by: 
\beq
  \langle \Sigma_z\rangle_d & =& \frac{ V_-^2\sin^2\theta - (V_+ + V_-\cos\theta)^2}{ 1+\beta_e\cos\theta} \nonumber\\
     \langle \Sigma_x\rangle_d &=  & \frac{ -2\sin\theta\, V_- (V_+ + V_-\cos\theta)}{ 1+\beta_e\cos\theta},\nonumber\\
  \langle \Sigma_z\rangle_u &= & \frac{  (V_+ - V_-\cos\theta)^2 - V_-^2\sin^2\theta}{ 1-\beta_e\cos\theta}  \,\,\,{\rm and}\nonumber\\
     \langle \Sigma_x\rangle_u &=  & \frac{ -2\sin\theta\, V_- (V_+ - V_-\cos\theta)}{ 1-\beta_e\cos\theta}.
 \eeq
In the fully relativistic limit, $V_+=V_-=1/\sqrt2$, the spin in either state points in the negative $\hat\theta$ direction,  owing to the state having negative chirality.    On the other hand, in the non-relativistic limit, $V_+=1, V_-=0$, the spin in either state lies along the z axis, independent of $\theta$, i.e., 
$\langle \Sigma_z\rangle_d =  -1 = -  \langle \Sigma_z\rangle_u$;  the spin of a non-relativistic electron ($\beta_e\to 0$) is not rotated in the weak interaction. 

\begin{figure}[t]
\includegraphics*[width=\linewidth]{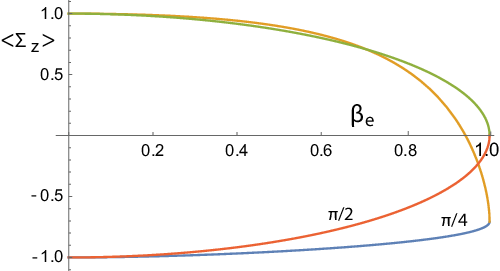}
\caption{Spin projection along $\hat z$ for an electron emerging at $\theta = \pi/2$  in the state $\Psi_{eu}$  (upper branch, green) and  $\Psi_{ed}$  (lower branch, red), and $\theta = \pi/4$  in $\Psi_{eu}$  (upper branch, yellow) and  $\Psi_{ed}$  (lower branch, blue) vs. $\beta_e$ from 0 to 1, showing the evolution from no 
spin bending in the non-relativistic limit to negative helicity in the fully relativistic limit.  This result applies for both inverse beta decay and s-wave scattering, as discussed in the text.    Essentially, the spin direction along $\hat\theta$ rotates
clockwise in the $(z,x)$ plane from its original state long $\hat z$ at small $\beta_e$, to negative helicity in the very relativistic limit. At 
$\pi/4$, the spin of an electron in  $\Psi_{eu}$ points in the negative z direction for $\beta_e$ above $\simeq 0.94$.
}
\label{90-45degrees}
\end{figure}

   Figure~\ref{90-45degrees} shows how the spin projection along z evolves with increasing $\beta_e$ from 0 to 1, at angles $\theta = \pi/2$ and $\pi/4$, for both states $\Psi_{ed}$ and $\Psi_{eu}$.    At $\theta=0$, $\langle\Sigma_z\rangle_d = -1 =  -\langle\Sigma_z\rangle_u$, regardless of $\beta_e$, while at $\theta=\pi/2$, $\langle\Sigma_z\rangle_d = -\sqrt{1-\beta_e^2} =  -\langle\Sigma_z\rangle_u$, and $\langle\Sigma_x\rangle_d =\langle\Sigma_x\rangle_u= -\beta_e$.      The same results hold for s-wave scattering of a spin-1/2 fermion of finite mass.
 As we see, the correlation of helicity and direction is a relativistic effect.  Similarly the transversality condition on a scattered photon spin, discussed earlier, is a consequence of the photon being massless.
   
 The mean helicity of the emitted electron is
\beq
  \langle h \rangle = \langle \Sigma_z\rangle \cos \theta + \langle \Sigma_x\rangle \sin \theta.
\eeq  
For a left-handed incident neutrino,  
$\langle h \rangle_{\nu = L} = -(\cos\theta + \beta_e)/(1+\beta_e\cos\theta)$, and for a right-handed incident neutrino,
$\langle h \rangle_{\nu = R} = (\cos\theta - \beta_e)/(1-\beta_e\cos\theta)$.   Weighting the two terms by $(1\pm\beta_\nu)/2$ (where
 $\beta_\nu$ and $\beta_e$ are correlated) we find in toto,
\beq
   \langle h \rangle = -\frac{\beta_e \sin^2\theta + \beta_\nu(1-\beta_e^2)\cos\theta}{1-\beta_e^2\cos^2\theta}.
 \label{ehel}
\eeq

  The $\theta$ dependence of the mean helicity, Eq.~\eqref{ehel}, as well as the cross section, Eq.~\eqref{eang}, should, in principle, be detectable with incident laboratory or solar neutrinos \cite{cr51}.  The mean helicity varies from  $-\beta_\nu$ in the forward direction to $-\beta_e$ at 90$^\circ$.  In the search for relic neutrinos, however,  one will not see the $\cos\theta$ dependence, owing to their isotropy.

\section{Beta decay}
\label{betadecay}

  We turn next to the state of the electron and antineutrino produced in a beta decay process,
neglecting the recoil of the nucleus.  The electron and antineutrino are each produced in spherical states of total angular momentum 1/2, as is the electron in inverse beta decay;  the electron has negative and the antineutrino positive chirality. In a beta decay that does not change the angular momentum of the nucleus, e.g., a $0^+ \to 0^+$ transition,
the electron-antineutrino system is entangled in a spin-singlet state.   We take the spin quantization axis along an arbitrary direction, labelled $\hat z$; the rotational invariance of a spin-singlet implies that the total state is independent of the choice of axis.
 
  The wave functions of the positive chirality emitted right-handed or left-handed helicity antineutrino, respectively, have the form, 
\beq
  \Psi_{\bar\nu u}(\vec r\,) &= & \frac{h_0(p_\nu r)}{\sqrt2}  \big( W_+ +W_-\cos\theta_{\bar\nu}, W_-\sin\theta_{\bar\nu}  e^{i\phi_{\bar\nu}}, \nonumber\\ &&\hspace{30pt} W_+ +W_-\cos\theta_{\bar\nu}, W_-\sin\theta_{\bar\nu}  e^{i\phi_{\bar\nu}}\big)^T,\nonumber\\
    \Psi_{\bar\nu d}(\vec r\,) &=&\frac{h_0(p_{\bar\nu}r)}{\sqrt2}\big(W_-\sin\theta_{\bar\nu}  e^{i\phi_{\bar\nu}},W_+ - W_-\cos\theta_{\bar\nu},\nonumber\\&&\hspace{30pt} W_-\sin\theta_{\bar\nu}  e^{i\phi_{\bar\nu}},W_+ - W_-\cos\theta_{\bar\nu}\big)^T. \nonumber\\
\eeq
The angular momenta of the states $\Psi_{\bar\nu u}$ and $ \Psi_{\bar\nu d}$ are $(J,J_z) = (1/2,\pm1/2)$, respectively.
The final electron states are the same as in inverse beta decay,  Eqs.~\eqref{psied} and \eqref{psieu}, with the angles $\theta_e$ and $\phi_e$ of the electron.

    The combined wave function of the electron-antineutrino pair is then,
\beq
   \Psi_{e \bar\nu} = \frac1{\sqrt2}\big(\Psi_{e d}(r_e)\Psi_{\bar\nu u}(r_{\bar\nu}) -\Psi_{e u}(r_e)\Psi_{\bar\nu d}(r_{\bar\nu}) \big).
\eeq
The amplitudes in beta decay for the electron to emerge with helicity $h_e=L$ or $R$ and the antineutrino with right-handed helicity, at relative angle $\theta$ between the electron and antineutrino [$\cos \theta = \cos\theta_e\cos\theta_{\bar\nu} +\sin\theta_e\sin\theta_{\bar\nu}\cos(\phi_e-\phi_{\bar\nu}) $], are given in terms of the amplitudes in Eqs.~\eqref{aL} and \eqref{aR} (where $\theta$ there should now be read as the relative angle between the electron and antineutrino) by $\frac12\sqrt{1+\beta_{\bar\nu}}\,a^{(d)}_{h_e}$;  similarly the amplitude for the antineutrino to have left-handed helicity is $-\frac12\sqrt{1-\beta_{\bar\nu}}\,a^{(u)}_{h_e}$.

   The  angular dependence of the total cross section, or equivalently the electron angular distribution with respect to the neutrino direction, given by
\beq
\frac{d\sigma}{d\Omega_{e\bar\nu}}&\sim& (1+\beta_{\bar\nu})|\Psi_{ed}|^2 + (1-\beta_{\bar\nu})|\Psi_{eu}|^2 \nonumber\\
 &\sim& 1+\beta_e\beta_{\bar\nu}\cos\theta,
\eeq
is the same as in inverse beta decay, Eq.~\eqref{eang}, and is thus independent of whether a neutrino is incident or an antineutrino is emitted.  The cross section  vanishes, as expected, if the electron and antineutrino are massless and emerge back to back.
Similarly the mean electron helicity is given by Eq.~\eqref{ehel}.

    \section{Lepton-antilepton annihilation}
\label{pairs}

    We return to the question of angular momentum conservation in lepton pair annihilation, considering first the process, $\bar\nu_e +  \nu_e \to \bar\nu_\mu +  \nu_\mu$, assuming all the neutrinos to be massless.  With the  $\bar\nu_e$ travelling in the positive z direction and the $\nu_e$ in the negative z direction, the initial state of the system has angular momentum $(J,J_z) = (1,1)$ with $J_z$ measured along the positive z direction.    The relative wave function of the final state can be written as the product of spherical waves for the antineutrino in the $(\theta,\phi)$ direction and the neutrino in the $(\pi-\theta, \phi+\pi)$ direction:
 \beq
  \Psi_{\nu_\mu \bar\nu_\mu} =  \Psi_{\nu_\mu u}(\pi-\theta,\phi+\pi)\Psi_{\bar\nu_\mu u}(\theta,\phi),
  \label{pairwf}
\eeq
a product of $ \gamma_5=-1$ and $\gamma_5=+1$ states.   The spherical waves for massless neutrinos are,
\beq
  \Psi_{\bar\nu_\mu u}(\vec r\,) &= & \frac{h_0(p_\nu r)}{2}  \big( 1 +\cos\theta, \sin\theta e^{i\phi},\nonumber\\ && \hspace{42pt}
   1 +\cos\theta, \sin\theta  e^{i\phi }\big)^T \,\,\,{\rm and}\nonumber\\
\Psi_{\nu_\mu u}(\vec r\,)&=&\frac{h_0(p_\nu r)}{2}\big(1-\cos\theta,  \sin\theta  e^{i\phi},  \nonumber\\ && \hspace{24pt} 
  -( 1 -\cos\theta),  -\sin\theta  e^{i\phi} \big)^T ;
\eeq
thus,
\beq
  \Psi_{\nu_\mu u}(\pi-\theta,\phi+\pi)&=&\frac{h_0(p_\nu r)}{2}\big(1+\cos\theta,  -\sin\theta  e^{i\phi},  \nonumber\\ && \hspace{2pt}   
     -( 1 +\cos\theta),  \sin\theta  e^{i\phi} \big)^T. 
\eeq
That $J_z=1$ is easily verified, as before,  by acting with $S_{z\bar\nu_\mu}+S_{z\nu_\mu} +L_z$.   

  The cross section for observing the antineutrino at angle $\theta$ is then 
\beq
   \frac{d\sigma}{d\Omega} \sim  |\Psi_{\nu_\mu \bar\nu_\mu}|^2 \sim (1+\cos\theta)^2.
   \label{nunununu}
\eeq
As with the electron in inverse beta decay, detection of the final muon antineutrino along a given direction projects the total spin along that direction.

      Angular momentum conservation in the electromagnetic annihilation process \cite{epair,peskin,bartel,adeva} $e^+e^- \to \mu^+\mu^-$ is similar.  We assume that the positron is incident along the +z axis and the electron is incident along the -z axis. The incident $e^+e^-$ pair must have total spin one to produce the required intermediate photon, the total spin projection along the z axis must be $\pm 1$.  
We assume, for the sake of the argument, that the positron is right spin-polarized and the electron is left spin-polarized.  
Thus the initial angular momentum is $(J,J_z)=(1,1)$.

    In the limit in which the center of mass energy is much larger than the muon rest mass, which we consider, the electromagnetic interaction, of the form $\bar u V u = \bar u_L V u_L + \bar u_R V u_R$, leads to either a left-handed helicity $\mu^-$ with a right-handed helicity $\mu^+$, or a right-handed helicity $\mu^-$ with a left-handed helicity $\mu^+$.   Thus the final state has a spherical structure similar to Eq.~\eqref{pairwf},
\beq
  \Psi_{\mu^-\mu^+} &=&  \Psi_{\mu^- u}(\pi-\theta,\phi+\pi)\Psi_{\mu^+ u}(\theta,\phi) \nonumber\\  &&+ \Psi_{\mu^+u}(\pi-\theta,\phi+\pi)\Psi_{\mu^- u}(\theta,\phi). 
\eeq
The two terms are orthogonal and so their relative phase plays no role here.    The differential cross section is given by
\beq
  \hspace{-6pt}\frac{d\sigma}{d\Omega}  &\sim&  |\Psi_{\mu^-\mu^+}|^2 \nonumber\\ &\sim& (1+\cos\theta)^2 +  (1-\cos\theta)^2  \sim 1+\cos^2\theta,
\eeq
the well known result.

    One usually imagines that the muons emerge along the directions $(\theta,\phi)$ and $(\pi-\theta,\pi+\phi)$ with respect to the incident electron-positron, and with spin polarizations along these directions.   Since the individual spherical waves each have angular momentum 1/2 along the positive z direction,  this picture runs again into the same question of how angular momentum is conserved in a process in which the total spin state starts as  $|1,1\rangle_0$ but is measured to be $|1,\pm 1\rangle_\theta$, rotated by angle $\theta$.

\section{Conclusion}
\label{conclusion}

    Our original question was how angular momentum is conserved in interactions where the initial angular momentum is in a given direction, and the final angular momentum, once the direction of the final state particle (or particles) is
determined, is in a different direction.    The key to understanding angular momentum in the various processes we considered -- elastic photon scattering on an atom,  scattering of massless spin-1/2 particles, inverse beta decay, beta decay, neutrino-antineutrino annihilation, and electron-positron  annihilation to $\mu^+\mu^-$ -- is to work in terms of the spherical states  for the particles in question, which have the same angular momentum as the initial state.  For spin-1/2 particles the spherical states are spin-1/2 Dirac states.   We have shown that prior to measurement of the final particle, angular momentum is indeed conserved in all these processes. The apparent rotation of the angular momentum direction takes place in the measurement, where using localized devices, one cannot detect the angular momentum transferred to a realistic detector that is sufficiently localized to allow a direction measurement.   Conservation of angular momentum in these situations can only be accounted for by taking into account the angular momentum transferred to the detector.   
   
   The spherical wave function of the final states considered here is an entangled state of the spin and momentum directions.
A Dirac spherical wave with angular momentum along the $\pm$ z direction has the remarkable property that it is a linear superposition of momentum states with spin $\pm$ 1/2 along the momentum direction, a hedgehog type solution.     We have shown that a host of results on the angular and helicity distributions of particles produced in the weak interaction  processes can be derived in terms of these spherical spin-1/2 states.  This approach is readily applicable to other processes, e.g.,  parity violation in lepton pair annihilation, electron-positron annihilation into two photons, pion decay, related pair creation process in peripheral ultrarelativistic heavy ion collisions, and understanding the quantum mechanics of heavy quark pair, e.g., top--anti-top, production in colliders \cite{atlas,cms,tao}.

\acknowledgments

   Author GB is grateful to Tony Leggett, David Mermin, Wolfgang Ketterle, and Peter Abbamonte for very insightful discussions on
quantum measurement, to Tetsuo Hatsuda, Naoki Yamamoto,  Konstantin Bliokh, and Franco Nori for illuminating discussions of the spin-Hall effect during GB's stay in RIKEN iTHEMS, and to Xiangdong Ji for helpful comments.  This research was supported in part by National Science Foundation grant PHY-2111046, and by the Japan Science and Technology Agency (JST) as part of the Adopting Sustainable Partnerships for Innovative Research Ecosystem (ASPIRE), Grant Number JPMJAP2318, and was carried out in part at the Aspen Center for Physics, which is supported by NSF grant PHY-2210452.  Nordita is supported in part by NordForsk.

\end{document}